\documentstyle[preprint,aps]{revtex}
\newcommand{\beq}{\begin{equation}}
\newcommand{\eeq}{\end{equation}}
\newcommand{\beqa}{\begin{eqnarray}}
\newcommand{\eeqa}{\end{eqnarray}}
\newcommand{\ba}{\begin{array}}
\newcommand{\ea}{\end{array}}

\begin{document}
\draft

%@twocolumnfalse\endcsname
%%%%%%%%%%%%%%%%%%

\widetext 

\title{Resonances and Chaos in the Collective Oscillations 
of a Trapped Bose Condensate} 
\author{Luca Salasnich} 
\address{Istituto Nazionale per la Fisica della Materia, Unit\`a di Milano,\\ 
Dipartimento di Fisica, Universit\`a di Milano, \\ 
Via Celoria 16, 20133 Milano, Italy} 

\maketitle

\begin{abstract} 
We investigate the collective oscillations of a Bose condensate 
in an anisotropic harmonic trap. We analytically calculate 
the values of the trap deformation for which the normal 
modes are in resonance. In presence of low-order resonances, 
the amplitude dependence of frequencies is strongly enhanced. 
We also study the transition to chaos 
of the collective modes as a function of the trap anisotropy. 
\end{abstract}

\vskip 1.cm


%%%%%%%%%%%%%%%%%%
%]
%%%%%%%%%%%%%%%%%%

\narrowtext

The Bose-Einstein condensation (BEC) 
is the macroscopic occupation of the single-particle ground-state of a system 
of bosons. In the 1995 BEC has been achieved with dilute vapors 
of confined alkali-metal atoms.$^{1-3}$ 
The experiments with alkali-metal atoms generally consist 
of a laser cooling and confinement in an external potential 
(magnetic or magneto-optical trap) and an evaporative cooling.$^{1-3}$ 
Nowadays a dozen of experimental groups have obtained Bose condensates 
by using different atomic species and geometries of the confining trap. 
\par 
The theoretically predicted low-energy collective oscillations 
of the Bose condensate$^{4}$ have been experimentally confirmed by laser 
imaging techniques.$^{5}$  Moreover, the dynamics of the 
collective oscillations of the condensate has been 
theoretically studied beyond the linear regime, showing strong 
enhancement of the amplitude dependence of frequencies 
in presence of resonances.$^{6}$ 
\par
In this paper we derive analytical formulas that give 
the values of trap anisotropy for which the normal modes 
of the collective oscillations are in resonance. We discuss 
also the transition to chaos of the collective modes 
by increasing the energy 
of the system, namely by considering oscillations of very 
large amplitude. Note that the chaoticity of the 
high-energy single-particle excitations of the condensate 
has been shown in the classical limit 
by Fliesser {\it et al.}$^{7}$ 
Here, we consider instead the low-energy 
collective excitations: they have 
a classical-like behavior but satisfy the quantum mechanical 
equations of a superfluid in an external potential.$^{4}$ 
\par 
In current experiments alkali vapors are quite dilute 
and at zero temperature the atoms are practically 
all in the condensate.$^{1-3}$ 
The dynamics of the Bose condensate 
can be accurately described by the Gross-Pitaevskii (GP) 
equation$^{8}$ of mean-field approximation 
\beq 
i\hbar {\partial \over \partial t} \phi({\bf r},t) 
= \left[ -{\hbar^2 \over 2 m} \nabla^2 + U({\bf r}) 
+ g |\phi({\bf r},t)|^2 \right] \phi({\bf r},t) \; ,  
\eeq 
where $\phi ({\bf r},t)$ is the macroscopic wavefunction (order parameter) 
of the condensate, normalized to the number $N$ of condensed atoms. 
The external trap is well approximated by an anisotropic 
harmonic potential 
\beq 
U({\bf r})={m\over 2} (\omega_1 x^2 + \omega_2 y^2 + \omega_3 z^2 ) \; . 
\eeq
and the parameter of the nonlinear term $g={4\pi \hbar^2 a_s/m}$ 
is the scattering amplitude with $a_s$ the s-wave scattering length. 
It is useful to introduce the geometric average 
$\omega_h=(\omega_1 \omega_2 \omega_3)^{1/3}$ of the trap frequencies 
and the characteristic length $a_h=(\hbar/m\omega_h)^{1/2}$ of the trap. 
In previous papers we have studied the properties of the Bose condensate 
with attractive interaction, in particular for $^7$Li atoms.$^{9,10}$ 
Here, instead, we consider only the case of a repulsive interaction. 
\par 
The complex wavefunction $\phi({\bf r},t)$ 
of the condensate can be written in terms of 
a modulus and a phase, as follows 
\beq
\phi({\bf r},t) = \sqrt{ \rho({\bf r},t)} \; e^{iS({\bf r},t)} \; . 
\eeq
The phase $S$ fixes the velocity field ${\bf v}=(\hbar /m)\nabla S$. 
The GP equation can hence be rewritten in the form of two coupled 
hydrodynamic equations for the density and the velocity field.$^{4}$ 
If the repulsive interaction among atoms is strong enough, 
i.e. $Na_s/a_h>>1$, then one can safely neglect the kinetic 
pressure term in the equation for the velocity field. 
In this way, the hydrodynamic equations read 
$$
{\partial \over \partial t} \rho + \nabla \cdot ({\bf v} \rho) = 0 \; ,
$$
\beq 
m {\partial \over \partial t} {\bf v} + \nabla \left( U + g \rho  + 
{mv^2\over 2} \right) =0 \; . 
\eeq 
These equations are those of a superfluid in an external 
potential.$^{4}$ 
The ground-state solution (${\bf v}=0$) of the hydrodynamic 
equations is $\rho({\bf r})  = g^{-1} \left[ \mu - U({\bf r}) \right]$, 
in the region where $\mu > U({\bf r})$, and $\rho = 0$ outside. The 
normalization condition on $\rho({\bf r})$ provides 
$\mu = (\hbar \omega_h/2)(15Na_s/a_h)^{2/5}$. 
\par
An analytic class of solutions of Eq. (4) 
has been found found by Dalfovo {\it et al.}$^{6}$ 
writing the density in the form 
$\rho({\bf r},t) = a_0(t) - a_1(t)x^2 - a_2(t)y^2 -a_3(t)z^2$ 
in the region where $\rho({\bf r},t)$ is positive, and the 
velocity field as 
${\bf v}({\bf r},t) = {1\over 2}\nabla 
[ b_1(t) x^2 + b_2(t) y^2 + b_3(t) z^2 ]$. 
The coefficient $a_0$ is fixed by the normalization of the 
density $a_0=(15N/8\pi)^{2/5}(a_1 a_2 a_3)^{1/5}$. 
By inserting these expressions in the hydrodynamic equations 
one finds 6 coupled differential equations for the time-dependent 
parameters $a_i(t)$ and $b_i(t)$. 
By introducing new variables $q_i$, defined by $a_i=m\omega_i^2 
(2gq_i^2 q_1 q_2 q_3)^{-1}$, the hydrodynamic equations give 
$b_i = {\dot q_i}/q_i$ and 
\beq
{\ddot q_i} + \omega_i^2 q_i = 
{\omega_i^2 \over q_i q_1 q_2 q_3} \; ,
\eeq
with $i=1,2,3$. 
It is important to observe that, using the new variables $q_i$, the 
equations of motion do not depend on the value of the coupling 
constant $g$. In terms of $q_i$ the mean square radii of the condensate 
are $<x_i^2>=(2\mu/m\omega_i)q_i^2$ 
and the velocities are $<v_i^2>=(2\mu/m\omega_i){\dot q}_i^2$ 
(see also Ref. 6). 
\par
The Eq. (5) are the classical 
equations of motion of a system with coordinates $q_i$ and 
Lagrangian given by 
\beq
L={1\over 2} \sum_{i=1}^3(\omega_i^{-2} {\dot q}_i^2 - q_i^2 ) 
- {1\over q_1 q_2 q_3} \; . 
\eeq 
This Lagrangian describes the 
collective modes of the Bose condensate for $N a_s/a_h >>1$.  
In such a case the collective dynamics of the condensate 
does not depend on the number of atoms 
or the scattering length. 
The low-energy collective excitations of the condensate are 
the small oscillations of variables $q_i$'s around the equilibrium 
point, i.e. the minimum of the effective potential 
$\sum_{i=1}^3 q_i^2 + 1/(q_1 q_2 q_3)$. 
The normal mode frequencies $\Omega$ 
of the condensate are given by the following equation  
\beq
\Omega^6
-3\left(\omega_1^2+\omega_2^2+\omega_3^2\right)\Omega^4+
8\left(\omega_1^2\; \omega_2^2+\omega_1^2\; \omega_3^2 
+\omega_2^2\; \omega_3^2 \right)\Omega^2 
-20\; \omega_1^2\; \omega_2^2\; \omega_3^2=0 \; .
\eeq   
Note that this formula has been recently obtained by using 
a variational approach$^{11}$  and also by studying 
hydrodynamic density fluctuations of the condensate.$^{12}$ 
For an axially symmetric trap, where 
$\omega_1=\omega_2=\omega_{\bot}$, the previous equation gives  
$$
\Omega_{1,2}=\sqrt{2 + {3\over 2}\lambda^2 \pm 
{1\over 2}\sqrt{16 + 9\lambda^4 - 16\lambda^2} } 
\; \omega_{\bot} \; ,
$$
\beq
\Omega_3 =\sqrt{2} \; \omega_{\bot} \; ,
\eeq 
where $\lambda = \omega_3/\omega_{\bot}$ is the anisotropy 
parameter of the trap. The frequencies $\Omega_{1,2}$ are 
the $m_z=0$ monopole and quadrupole modes, where $m_z$ is the third 
component of the angular momentum. Instead the frequency 
$\Omega_3$ is the $m_z=2$ quadrupole mode.$^{4}$ 
It is important to observe that the experimental results obtained 
with sodium vapors at MIT ($\lambda = \sqrt{8}$) 
are in good agreement with these theoretical values.$^{13}$ 
\par
The normal modes are functions of the trap frequencies, 
i.e. of the trap deformation. Beyond the linear regime 
there is a strong enhancement of nonlinear effects 
when frequencies of different modes, 
or of their harmonics, coincide, i.e. when the normal 
modes are in resonance.$^{6,14}$ 
By using a perturbative approach, Dalfovo {\it et al.}$^{6}$ 
have identified 3 resonant values of the trap anisotropy. 
For such values, the numerical solution of Eq. (5) shows that 
the collective frequencies deviate from the 
predictions of Eq. (7) or (8) also for small amplitude oscillations. 
\par 
To find the BEC resonances we use a different and new strategy. 
In general, the resonance condition is given by 
\beq
n_1 \Omega_1 + n_2 \Omega_2 + n_3 \Omega_3  = 0 \; ,
\eeq 
where $(n_1,n_2,n_3) \in {\bf Z}^3\backslash \{{\bf 0}\}$ is a non-zero 
vector of integer numbers and $R=|n_1|+|n_2|+|n_3|$ is the order 
of the resonance. It is possible to numerically determine 
the values of the trap deformation for which the 
resonance condition is satisfied. 
Nevertheless, because in most experiments the confining trap 
has axial symmetry,$^{1-3}$ one obtains remarkable 
analytical results from Eq. (8) and (9) by imposing that 
one of the integer numbers $n_i$ is zero. 
In the case $n_1=0$, the resonance condition 
(between a $m_z=0$ mode and the $m_z=2$ mode) gives  
\beq 
\lambda = \sqrt{ 2 (2 n_2^2 - n_3^2) n_3^2 
\over (5 n_2^2 - 3 n_3^2) n_2^2 } \; .
\eeq 
In the case $n_2=0$, one has the previous formula with $n_1$ instead of 
$n_2$. Finally, in the case $n_3=0$ one finds 
\beq 
\lambda = {1\over 3} \sqrt{\left[5 \left({n_1^2\over n_2^2} 
+{n_2^2\over n_1^2}\right)-2\right]
\pm \sqrt{\left[5 \left({n_1^2\over n_2^2}
+{n_2^2\over n_1^2}\right) -2\right]^2 - 144}} \; . 
\eeq 
This formula gives the anisotropy parameter 
at the resonance condition between the two $m_z=0$ modes. 
\par
In Table 1 are shown the values of the trap anisotropy 
$\lambda$ for low-order resonances according to Eq. (10) and 
(11). Note that the values of $\lambda$ for the 
$(1,2,0)$ and $(0,1,2)$ resonances 
are that previously obtained by Dalfovo {\it et al.}.$^{6}$ 
As concerns resonances, 
the values of $\lambda$ for the experimental traps of MIT 
($\lambda = 0.077$) and JILA ($\lambda =\sqrt{8}$) are not 
of particular interest. Instead, it would be very interesting 
to study experimentally the system in the trap conditions 
given by Eq. (10) and (11). In fact, close to low-order resonances, 
even for relatively small amplitudes, 
the motion is rather complex and the dynamics 
can exhibit a chaotic-like behavior.$^{14}$ 
\par
One expects an order-chaos transition 
by increasing the amplitude of oscillations, thus by increasing 
the energy of the system with respect to the ground-state 
(minimum of the effective potential).$^{14}$ 
To check if such a scenario works 
in this context, we numerically study the chaotic 
dynamics of the $m_z=0$ modes 
(preliminar results can be found in Ref. 15). 
We put $q_1=q_2=q_{\bot}$ in the BEC lagrangian. 
By using the adimensional time 
$\tau = \omega_{\bot}t$, the BEC Hamiltonian of the $m_z=0$ 
collective modes reads 
\beq
H=p_{\bot}^2 + {1\over 2}\lambda^2 p_3^2 + q_{\bot}^2 
+ {1\over 2} q_3^2  + {1\over q_{\bot}^2 q_3} \; , 
\eeq
where $p_{\bot}=dq_{\bot}/d\tau$ and 
$p_3= \lambda^{-2}dq_3/d\tau$ are the conjugate momenta.$^{15}$  
We use a symplectic method to numerically compute 
the trajectories. The conservation 
of energy restricts any trajectory of the four-dimensional 
phase space to a three-dimensional energy shell. At a particular energy, 
the restriction $q_{\bot}=1$ defines a two-dimensional surface in 
the phase space, that is called Poincar\`e section. 
Each time a particular trajectory passes through the surface 
a point is plotted at the position of intersection $(q_3,p_3)$. 
We employ a first-order interpolation process to reduce inaccuracies 
due to the use of a finite step length. 
\par 
In Fig.1 and Fig. 2 we plot Poincar\`e sections of the system 
with $\lambda = 1.501$ (corresponding 
to the $(3,5,0)$ resonance) and $\lambda=\sqrt{8}$ (JILA experiment), 
respectively. In each panel there is a Poincar\`e 
section with a fixed value of the energy of the system. 
At each energy value, we have chosen different 
initial conditions [$q_{\bot}(0)$,$q_3(0)$,$p_{\bot}(0)$,$p_3(0)$] 
for the dynamics. Actually $p_{\bot}(0)$ has been fixed 
by the conservation of energy. Integration time is $400$ in 
adimensional units, that is less than $1$ second 
(the life-time of the condensate is about $10$ seconds). 
Chaotic regions on the Poincar\`e section 
are characterized by a set of randomly distributed points 
and regular regions by dotted or solid curves. 
\par
We use the adimensional parameter $\chi$, that is 
the relative increase of energy with respect to the ground-state. 
Fig. 1 shows that for $\lambda = 1.501$ there is 
an order-chaos transition by increasing $\chi$: 
a chaotic sea appears at low energy ($\chi=2/5$). 
On the contrary, Fig. 2 shows that for $\lambda =\sqrt{8}$ 
at $\chi =3/5$ the collective oscillations 
are inharmonic but still not chaotic. 
\par 
To estimate the chaoticity of the $m_z=0$ modes, 
we calculate the fraction of initial conditions that give 
rise to regular trajectories. In the ($\lambda$,$\chi$) plane 
we plot the configurations for which the $m_z=0$ oscillations are regular. 
The numerical results are shown in Fig. 3. 
As previously discussed, for $\lambda=1.501$ 
(or $\lambda$ very close to such a value) 
the system remains regular only at small $\chi$;  
for larger $\chi$ it becomes gradually chaotic. 
In general, the transition to chaos depends on $\chi$, but 
for $\lambda <<1$ and $\lambda >4$ the dynamics appears always regular. 
Actually, for $\lambda <<1$ the system is integrable because of 
an additional constant of motion.$^{16}$ 
The MIT trap ($\lambda=0.077$) belongs to this stable region. 
\par 
We have analyzed the Poincar\`e sections of the $m_z=0$ 
BEC Hamiltonian for all the resonant values of 
$\lambda$ that are shown in Tab. 1 (left). 
Among the resonances, the $(3,5,0)$ resonance 
($\lambda=0.888$ and $\lambda =1.501$)  
appears the most chaotic: this is the main prediction of our paper.  
\par 
An important question is the following: can BEC chaotic dynamics 
be experimentally detected? The answer is positive. 
It is important to observe that, strictly speaking, 
there is no dynamical chaos in quantum systems 
with a finite number of degrees of freedom 
(for a discussion of chaos and quantum chaos 
in field theories see Ref. 17). 
Chaotic behaviour is possible only as a transient with 
lifetime $t_H$, the so-called Heisenberg time.$^{18}$ 
Nevertheless, $t_H$ grows exponentially with the number of degrees of freedom 
and consequently the transient chaotic dynamics of quantum states 
and observables can be experimentally observed in Bose condensates.$^{19}$ 
Nowadays non-destructive images of the dynamics of the condensate 
can be obtained. The radius of the condensate 
as a function of time can be detected and its time evolution 
analyzed with great accuracy. Various initial conditions of the collective 
dynamics can be obtained by using laser beams or 
by modulating for a short period the magnetic fields which 
confine the condensate. 
\par
In conclusion, we have studied the collective oscillations of a 
Bose condensate in an anisotropic harmonic trap. 
We have found the values of trap deformation for which the normal 
modes of the collective oscillations are in resonance. 
For such values, nonlinear effects, like frequency shift and 
irregular patterns in the time evolution of the condensate shape, 
are strongly enhanced. 
We have analyzed the order-chaos transition of the $m_z=0$ 
collective oscillations by increasing the energy, 
namely the amplitude of oscillation. 
We have obtained compelling evidence that, 
for certain values of trap anisotropy, chaos is reached 
at very low energy. 
\par 
We hope that our calculations may stimulate 
new measurements on the collective oscillations of Bose condensates 
in anisotropic traps.  

\vskip 0.5 truecm 
\par
The author thanks Prof. L. Reatto for useful suggestions. 

\newpage 

\section*{References}

\begin{description}

\item{\ 1.} M.H. Anderson, J.R. Ensher, M.R. Matthews, C.E. Wieman, 
and E.A. Cornell, Science {\bf 269}, 189 (1995). 

\item{\ 2.} K.B. Davis, M.O. Mewes, M.R. Andrews, N.J. van Druten, 
D.S. Drufee, D.M. Stamper-Kurn, and W. Ketterle, Phys. Rev. Lett. 
{\bf 75}, 3969 (1995).

\item{\ 3.} C.C. Bradley, C.A. Sackett, J.J. Tollet, and R.G. Hulet, 
Phys. Rev. Lett. {\bf 75}, 1687 (1995). 

\item{\ 4.} S. Stringari, Phys. Rev. Lett. {\bf 77}, 2360 (1996). 

\item{\ 5.} D.S. Jin, J.R. Ensher, M.R. Matthews, 
C.E. Wieman, and E.A. Cornell, 
Phys. Rev. Lett. {\bf 77}, 420 (1996); 
R. Onofrio, D.D. Durfee, C. Raman, M. Kohl, C.E. Kuklewicz, 
W. Ketterle, preprint cond-mat/9908340. 

\item{\ 6.} F. Dalfovo, C. Minniti, S. Stringari, and 
L. Pitaevskii, Phys. Lett. A {\bf 227}, 259 (1997); 
F. Dalfovo, C. Minniti, and L. Pitaevskii, 
Phys. Rev. A {\bf 56}, 4855 (1997). 

\item{\ 7.} M. Fliesser, A. Csordas, P. Szepfalusy, and 
R. Graham, Phys. Rev. A {\bf 56}, R2533 (1997). 

\item{\ 8.} E.P. Gross, Nuovo Cimento 20(1961) 454; 
J. Math. Phys. {\bf 4}, 195 (1963); 
L.P. Pitaevskii, Zh. Eksp. Teor. Fiz. {\bf 40}, 646 (1961) 
[Sov. Phys. JETP {\bf 13}, 451 (1961)].

\item{\ 9.} A. Parola, L. Salasnich, and L. Reatto, 
Phys. Rev. A {\bf 57}, R3180 (1998); 
L. Reatto, A. Parola, and L. Salasnich, 
J. Low Temp. Phys. {\bf 113}, 195 (1998). 

\item{\ 10.} L. Salasnich, Mod. Phys. Lett. B {\bf 11} 1249 (1997); 
L. Salasnich, Mod. Phys. Lett. B {\bf 12}, 649 (1998); L. Salasnich, 
Phys. Rev. A {\bf 61}, 015601 (2000). 

\item{\ 11.} E. Cerboneschi, R. Mannella, E. Arimondo, and 
L. Salasnich, Phys. Lett. A {\bf 249}, 245 (1998). 

\item{\ 12.} S. Stringari, Phys. Rev. A {\bf 58} 2385 (1998). 

\item{\ 13.} D.M. Stamper-Kurn, H.J. Miesner, S. Inouye, 
M.R. Andrews, and W. Ketterle, Phys. Rev. Lett. {\bf 81}, 500 (1998).

\item{\ 14.} A.J. Lichtenberg and M.A. Lieberman, 
{\it Regular and Stochastic Motion} (Springer, New York, 1983); 
V.I. Arnold, {\it Mathematical Methods of Classical Mechanics} 
(Moscow, Nauka, 1974). 

\item{\ 15.} L. Salasnich, preprint chao-dyn/9906034, 
to appear in Progr. Theor. Phys. Suppl., No. {\bf 139} (2000). 

\item{\ 16.} L.P. Pitaevskii, Phys. Lett. A {\bf 221}, 14 (1996); 
L.P. Pitaevskii and A. Rosch, preprint cond-mat/9608135.  

\item{\ 17.} L. Salasnich, J. Math. Phys. {\bf 40}, 4429 (1999). 

\item{\ 18.} V.R. Manfredi and L. Salasnich, in 
{\it Perspectives on Theoretical Nuclear Physics VII}, pp. 319-324, 
Ed. A. Fabrocini {\it et al.} (Edizioni ETS, Pisa, 1999). 

\item{\ 19.} V.R. Manfredi and L. Salasnich, 
Int. J. Mod. Phys. B {\bf 13}, 2343 (1999). 

\end{description}

\newpage

\begin{center}
\begin{tabular} {|ccc||cc|}
\hline
Resonance & $\lambda_1$ & $\lambda_2$ & Resonance & $\lambda$ \\
\hline 
$(1,2,0)$ & $0.683$ & $1.952$ & $(0,1,1)$ & $1$     \\
$(1,3,0)$ & $0.438$ & $3.081$ & $(0,1,2)$ & $1.512$ \\
$(1,4,0)$ & $0.320$ & $4.159$ & $(0,1,3)$ & $2.393$ \\
$(1,5,0)$ & $0.255$ & $5.226$ & $(0,2,1)$ & $0.454$ \\
$(2,5,0)$ & $0.527$ & $2.530$ & $(0,2,3)$ & $0.802$ \\
$(3,5,0)$ & $0.888$ & $1.501$ & $(0,3,1)$ & $0.300$ \\
\hline
\end{tabular}
\end{center}
{\bf Table 1}. Low-order resonances of the normal modes 
of the condensate and the corresponding values of the anisotropy 
parameter $\lambda$ of the external trap.  

\newpage

\section*{Figure Captions}

{\bf Figure 1}. Poincar\`e Sections of the $m_z=0$ modes. 
From left to right: 
$\chi=1/5$, $2/5$ and $3/5$, respectively. 
$\chi$ is the relative increase of energy with respect to the ground-state. 
Trap anisotropy $\lambda=1.501$. 
\\
\vskip 0.5 truecm 
{\bf Figure 2}. The same as in Figure 1. 
Trap anisotropy $\lambda=\sqrt{8}$ (JILA). 
\\
\vskip 0.5 truecm 
{\bf Figure 3}. Configurations of the ($\lambda$,$\chi$) plane 
for which the $m_z=0$ oscillations are regular. 
$\lambda$ is the trap anisotropy 
and $\chi$ is the relative increase of energy with respect 
to the ground-state. 
Full triangles: MIT trap. Full circles: $(3,5,0)$ resonance. 
Full squares: JILA trap. Open circles: all the other configurations. 

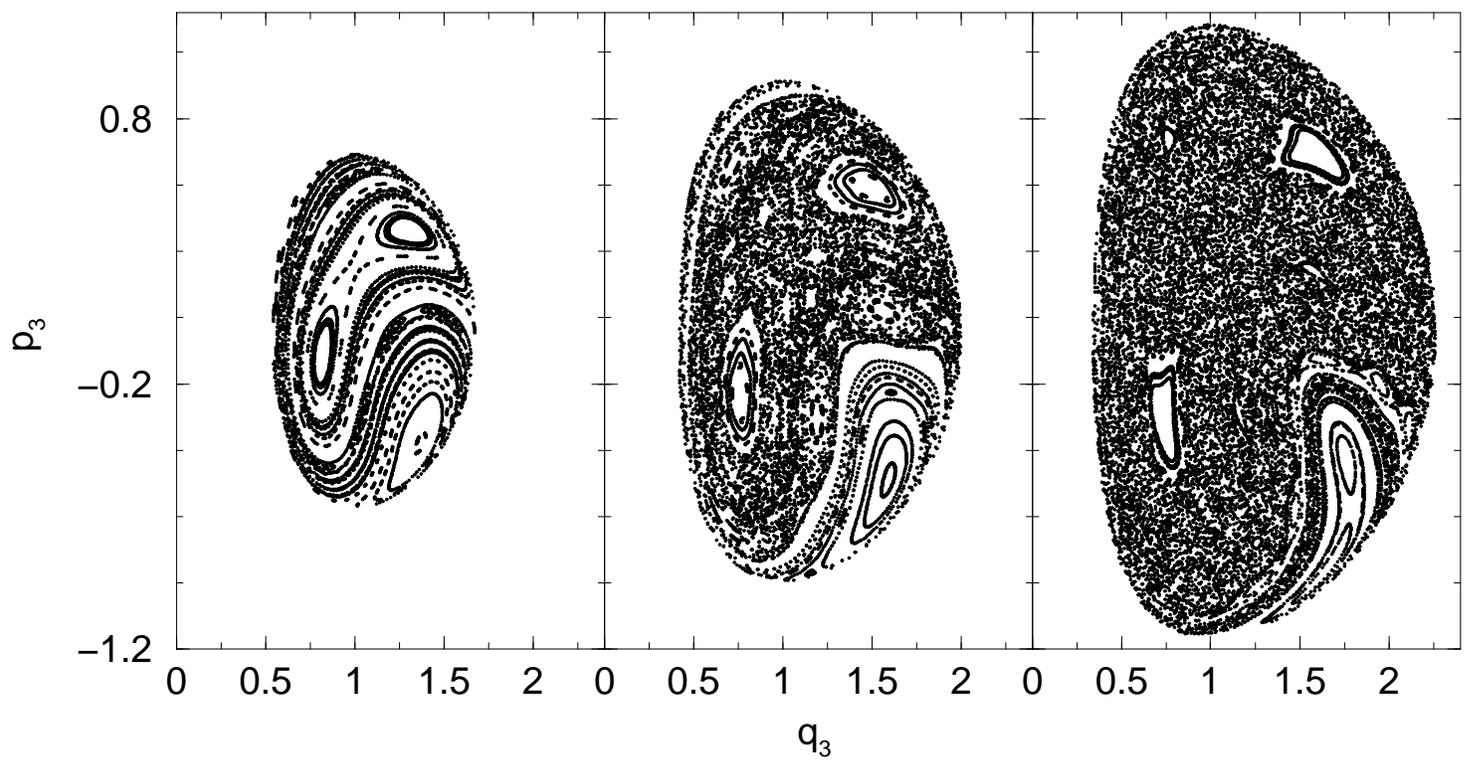
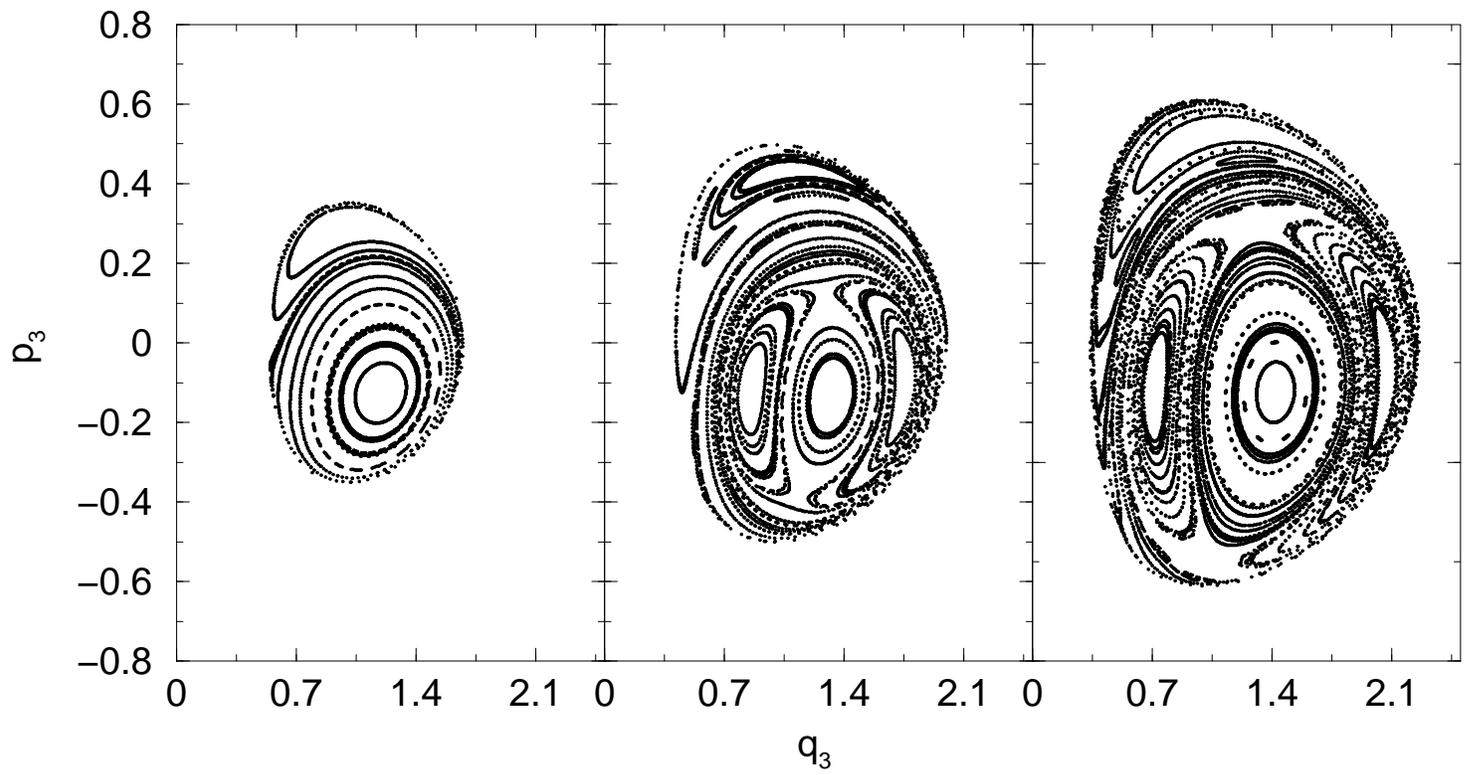
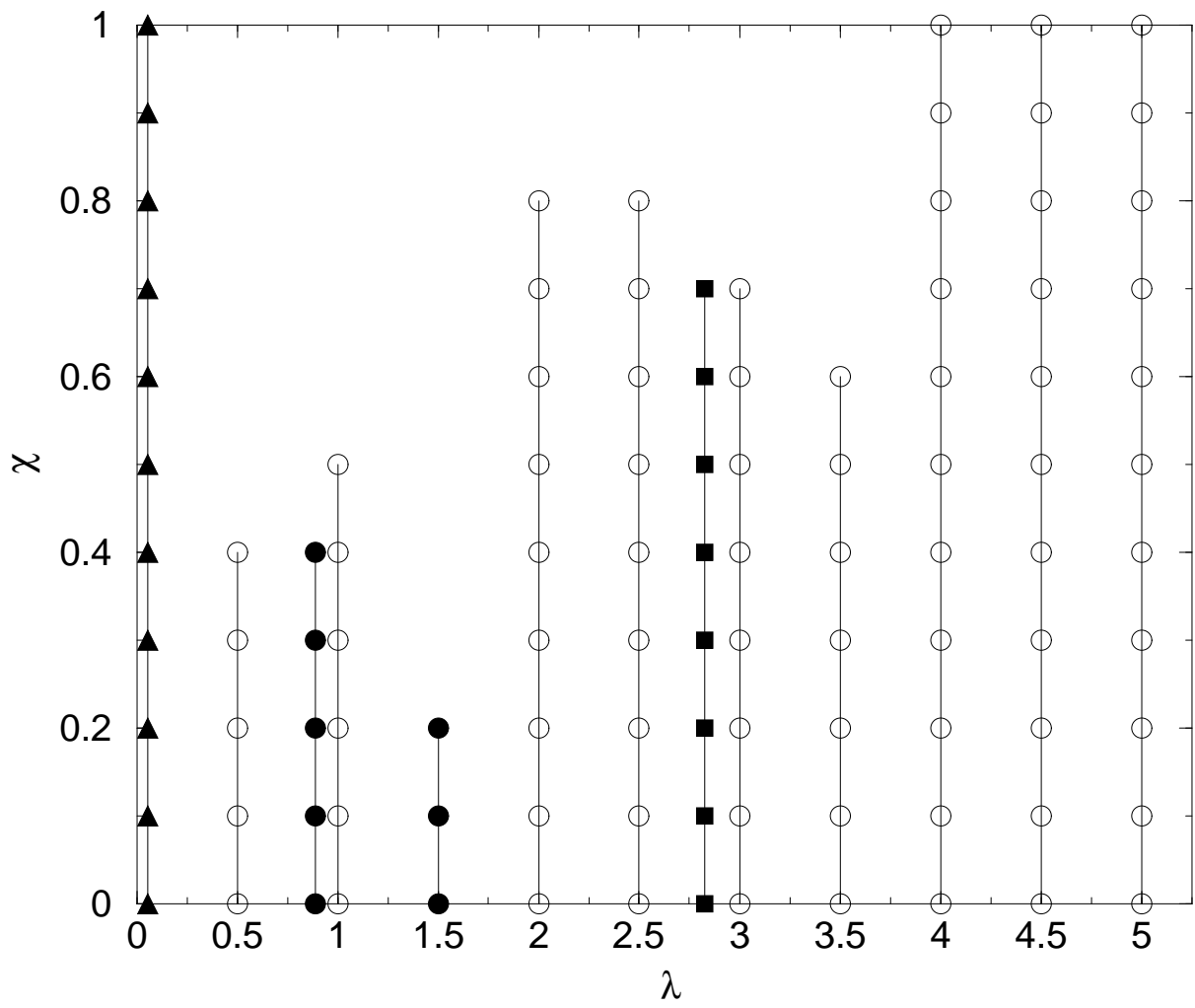
\end{document}